\documentclass[journal=jacsat,manuscript=article]{achemso}

\usepackage{chemformula} 
\usepackage[T1]{fontenc} 
\usepackage{graphicx} 
\usepackage{tabularx, makecell, booktabs}
\usepackage{amssymb} 
\usepackage{xcolor}
\newcommand{\fla}[1]{\begin{flalign}#1\end{flalign}}

\DeclareMathOperator{\sinc}{sinc}

\author{Maksim A. Smirnov}
\affiliation[KQC]
{Kazan Quantum Center, Kazan National Research Technical University named~after~A.N.~Tupolev, 420111, Kazan, Russia}

\author{Ilya V. Fedotov}
\affiliation[MSU]
{Lomonosov Moscow State University, 119992 Moscow, Russia}

\author{Anastasia M. Smirnova}
\affiliation[KQC]
{Kazan Quantum Center, Kazan National Research Technical University named~after~A.N.~Tupolev, 420111, Kazan, Russia}

\author{Albert F. Khairullin}
\affiliation[KQC]
{Kazan Quantum Center, Kazan National Research Technical University named~after~A.N.~Tupolev, 420111, Kazan, Russia}

\author{Andrei  B. Fedotov}
\affiliation[MSU]
{Lomonosov Moscow State University, 119992 Moscow, Russia}

\author{Sergey A. Moiseev}
\affiliation[KQC]
{Kazan Quantum Center, Kazan National Research Technical University named~after~A.N.~Tupolev, 420111, Kazan, Russia}

\email{s.a.moiseev@kazanqc.org}

\title{Bright ultra-broadband fiber-based biphoton source}

\abbreviations{IR,SFWM,SPDC, PCF}

\begin{document}

\begin{abstract}

In this Letter, we report a first experimental realization of  bright ultra-broadband (180 THz)  fiber-based biphoton source with widely spectrally separated signal and idler photons. 
Such a two-photon source is realized due to the joint use of broadband phase-matching of interacting light waves and high optical nonlinearity of a silica-core photonic crystal fiber.
The high performance of the developed fiber source identifies it as an important and useful tool for a wide range of optical quantum applications.
     
\end{abstract}

\section{Introduction}

Biphotons are a quantum state of light, consisting of correlated pairs of photons \cite{klyshko1967coherent, Harris_1967}. It is now a generally accepted fact that biphotons are useful and important tools for fundamental studies
\cite{Weihs1998,Howell2004}
and quantum optics applications, such as quantum communication \cite{Gisin_2002,Cozzolino_2019}, quantum imaging \cite{Moreau_2019}, spectroscopy 
and microscopy applications \cite{Dorfman_2018,Zhang2022}.
Special scientific interest is growing in biphotons with spectral band in the region of several tens of terahertz or more (ultra-broadband biphotons) \cite{Nasr_2008_PhysRevLett,chekhova2018broadband,Javid_PhysRevLett_2021,Katamadze_2022}.
Such bright biphoton sources can provide more efficient two-photon excitation and time-resolved  spectroscopy of molecular systems at low photon flux compared to conventional laser sources \cite{dayan2005nonlinear, lee2006_ETPA,Dorfman_2016,tabakaev2021_ETPA}.
In quantum optical coherence tomography (QOCT), ultra-broadband biphotons with individual spectral bands also promises achieving a submicron axial resolution \cite{nasr2003demonstration,Valles_QOCT_2018}.

\textcolor{black}{The creation of a biphoton source with an extremely wide spectral width and high generation efficiency is a challenging task for applied quantum optics. 
The broadband biphoton sources based on spontaneous parametric down-conversion (SPDC) can be created in  thin nonlinear crystals \cite{Katamadze_2015_Broadband_biphotons}, sub-wavelength nonlinear films~\cite{Chekhova_2021_entangled_thin},  aperiodically poled lithium niobate crystal~\cite{chekhova2018broadband,Sensarn_Harris_PhysRevLett_2010} or periodically poled lithium niobate waveguide ~\cite{KumarYadav_OptLett_2022, Javid_PhysRevLett_2021}.
Nowadays fiber biphoton sources are nearly as popular as crystal-based ones, especially sources based on spontaneous four-wave mixing (SFWM) in photonic crystal fibers (PCF)~\cite{Garay-Palmett_2023,knight2000anomalous,Smith_Walmsley_2009}  and tapered micro/nanofibers~\cite{kim2019photon,shukhin2020heralded}.}

The development of PCF-based sources has the following advantages: increasing coupling efficiency with fiber optic telecommunications; ability to control fiber dispersion behavior  through microstructure design.
There are two types of PCF, hollow core (HC-) and silica core (SC-), which can be used to generate extremely broadband biphotons. In the recent work~\cite{Chekhova_lopez2021fiber} has demonstrated biphoton generation using a HC-PCF with a total spectral width of 110~THz.
The possibility of creating broadband biphoton states with high degree of entanglement using SC-PCF has also been theoretically predicted (see works~\cite{Garay-Palmett_2023, petrov2019entropy}). 
The generation of biphotons in SC-PCF can provide an increase in the brightness of biphoton generation due to a higher value of the nonlinearity coefficient and a smaller mode area compared to gas-filled HC-PCF~\cite{PETROV_2019}. 
At the same time, the generation of broadband biphotons in SC-PCF  is accompanied by noise photons of spontaneous Raman scattering of the pump field
\cite{rarity2005photonic,Agrawal_2007_photon}. 
The noise photons limit the abilities of fiber-based sources and do not allow experimental generation of ultra-broadband biphotons.

In this letter,  we experimentally demonstrate the generation of biphoton fields with a record spectral width ($\approx$ 180 THz), high performance and low optical noise due to the use of SC-PCF with a silica core with a highly modified dispersion.
At the end of the article, we discuss the possibilities of using the proposed source of biphotons.

\textbf{Ultra-broadband phase-matching in PCF.} 
The generation of biphotons in the optical fiber is carried out by spontaneous four-wave mixing (SFWM) under the conditions of energy conservation   and phase matching 
\cite{Smith_Walmsley_2009,alibart2006photon}:
\begin{equation}
{2\omega_p=\omega_s+\omega_i},
\label{eq:Eq.1}
\end{equation}
\begin{equation}
2\beta_p = \beta_s+\beta_i+2{\gamma}P_p,
\label{eq:Eq.2}
\end{equation}
where $\omega_p$, $\omega_s$ and $\omega_i$ are the frequencies of the pump and generating photons (usually called signal and idler photons); $\beta_{j}=k_0(\omega_{j}) n_{eff}(\omega_{j})$ are propagation constants of fiber modes with frequencies $\omega_{j}$, where the first factor is the wave vector in vacuum, the second factor is the effective refractive index,
which depends on the structure of the PCFs, its geometric parameters and temperature,
$P_p$ is the pump power in the fiber, $\gamma$ is the effective nonlinear coefficient.

The biphotons generated at the fiber output are described by a two-photon state as \cite{Smith_Walmsley_2009}:
\begin{equation}
{|\Psi\rangle \propto  \iint d\omega_s\,d\omega_i\ F(\omega_s,\omega_i)\ \widehat{a_s}^+(\omega_s){ }\    \widehat{a_i}^+(\omega_i){ }\ |0_s\rangle\ |0_i\rangle},
\label{eq:Eq.3}
\end{equation}
where $F(\omega_s,\omega_i)$  is the joint spectral amplitude function (JSA), which is dependent on both the pump spectrum and the phase-matching $sinc$-function as \cite{migdall2013single}
\begin{equation}
{F(\omega_s,\omega_i) =  \alpha(\omega_s + \omega_i) ~ \sinc(\Delta\beta L/2)~\exp{(i\Delta\beta L/2)}}.
\label{eq:Eq.4}
\end{equation}
where $\Delta\beta~(\omega_s,\omega_i, \omega_p)=\beta_s+\beta_i+2{\gamma}P_p-2\beta_p$ is the phase-mismatch function, $L$ is is the fiber length and $\alpha(\omega_s + \omega_i)$  is the pump spectrum envelope function. 
The expression $I_{JS}\ (\omega_s,\omega_i)=|F(\omega_s,\omega_i)|^2$ related to the probability of generating biphotons with circular photon frequencies $\omega_s$ and $\omega_i$, respectively. 
In our experiments, the spectrum of generated photon pairs significantly exceeded the spectral width of the pump pulse $\Delta\lambda_p$. \textcolor{black}{In order to find the spectrum of signal and idler photons, we can represent Eq.~\ref{eq:Eq.4} as}

\fla{
 I(\omega_s) =  | \int \alpha(\omega_p) ~ \exp   \left(\frac{i\Delta\beta(\omega_s,2\omega_p-\omega_s, \omega_p) L}{2}\right)&
 \\ \nonumber
  & \sinc  	\left( \frac{\Delta\beta(\omega_s,2\omega_p-\omega_s, \omega_p) L}{2}\right)~d\omega_p 
 |^2,
 }
\label{eq:Eq.5}  

\textcolor{black}{where the argument of the mismatch function $\omega_i$ is replaced in accordance with Eq.~\ref{eq:Eq.1}. The spectrum for an idler photon is found similarly, in this case the argument $\omega_s$ is replaced.}

First of all,  to maximize the bandwidth of signal photons (as well as  idler photons), it is necessary to provide condition under which $\Delta \beta$ has an extremum (a saddle point) when the signal frequency $\omega_s$ changes (i.e. $\Delta \beta / \Delta\omega_s$ $|_{\omega s0} = 0$)  for a fixed pump frequency $\omega_p$ , as was shown in \cite{Vanselow:19} for SPDC-based biphoton source. 
The condition for the extremum is the group-velocity matching of the signal and idler photons. If $\Delta \beta$ is slightly deviated from 0 by shifting the pump frequency $\omega_p$, then in the parabolic approximation of $\Delta \beta$ (since the linear term in the Taylor expansion vanishes),  phase-matching occurs already at two wavelengths around the $\omega_{s0}$.
If the deviation of $\Delta \beta$ from 0 between these two points is small enough, the power produced between these points and slightly beyond them will be almost as high as with perfect phase-matching, further increasing the bandwidth. 
Secondly, in the silica-core PCFs as well as in standard optical fibers, biphoton generation is strongly hindered by the noise caused by Raman scattering radiation with the large spectral band width of $\Omega_R/2\pi \sim$ 40 THz \cite{rarity2005photonic,Agrawal_2007_photon}.
This also motivates searching the generation modes at which  the frequencies of biphotons will be strongly separated from the pump spectra ($\omega_{s0}-\omega_{p} \gg \Omega_R$ ). 

All these conditions (extremum of $\Delta \beta$ and large spectral separating) cannot be implemented in a silica-core PCF with a slightly biased dispersion as well as in a standard fiber, which usually characterized by one ZDW-value.
We propose the using of PCF with highly modified dispersion characterized by two values of ZDW, in broadband biphoton sources since this type of PCF can provide an extremum of $\Delta \beta$, as far as is known from numerical calculations of corresponding phase-matching curves \cite{Garay-Palmett_2023}.  
We experimentally demonstrate the generation mode of the ultra-broadband biphotons in the visible-near infrared optical range by using this type PCF.

\begin{figure}[h!]
\centering
{\includegraphics[width=\linewidth]{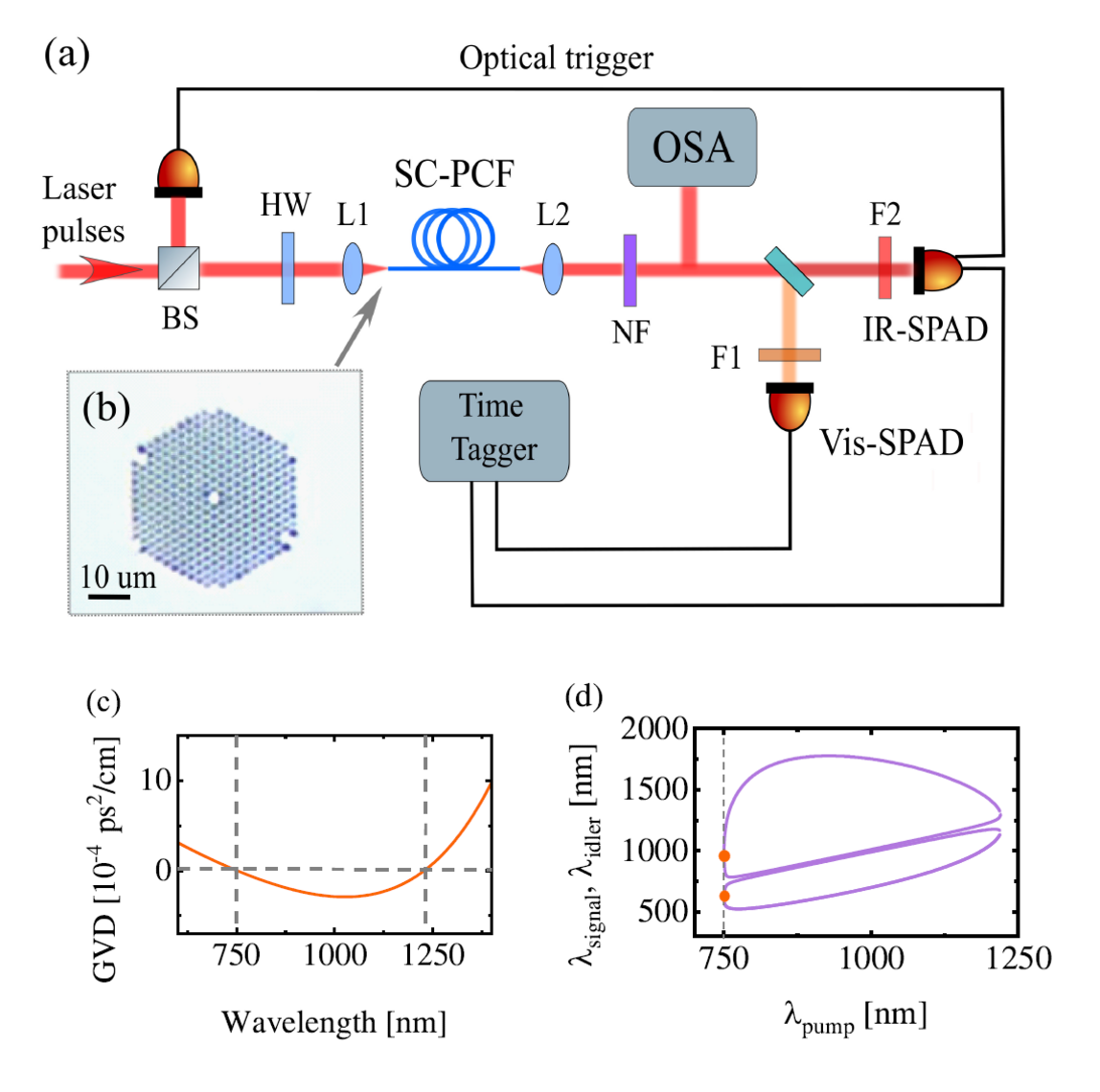}}
\caption{(a) The optical scheme of a fiber-based  biphoton source.
(b) Optical micrograph of SC-PCF used in experiments.
(c) Dispersion dependence of the  SC-PCF. 
(d) Phase-matching plot based on the dispersion. At a pump wavelength of 750 nm (dashed vertical line) the generation of photon pairs with ultrabroad spectral band at  the signal carrier wavelength 600 nm and the idler at 1000 nm is possible.}
\label{fig:1}
\end{figure}

\section{Results and discussion}

\textbf{Experimental methods.} 
Biphotons were generated by the SFWM method using femtosecond laser pump.
The experimental setup is shown in Fig.~\ref{fig:1}(a). The femtosecond pulses are produced by Ti:Sapphire laser with a central wavelength of 750 nm, pulse duration of $\sim 200$ fs, and repetition rate of $f_p=76$~MHz. 
In experiments, the width of the spectrum of laser pulses was controlled by limiting them on a homemade monochromator according to the work \cite{petrovnin2019broadband}.
The polarization direction of laser pulses was controlled by a half-wave plate ($HW$).
The pump beam is launched into the SC-PCF with an aspheric lens ($L1$) with numerical aperture $NA=0.5$. 
A spherical lens ($L2$) was  used for the output coupling. The pump power is controlled by a smoothly tunable neutral filter (not shown). 
After that, the pump was filtered by two notch filters ($NF$) with bandwidth of 34 nm.
The spectral characteristics of the generated  photons were measured using a spectrometer ($OSA$) based on a monochromator and a cooled CCD-matrix sensor (S7031-1006S, Hamamatsu).  

As an active medium for generating biphotons, we used a photonic crystal fiber "$NL-PM$ $750$" (NKT Photonics) with a length of $\approx100$ cm.
The optical micrograph of the SC-PCF cross-section with a silica-core diameter of $\sim$2~$\mu$m and a hole diameter of $\sim1$~$\mu$m is depicted in Fig. \ref{fig:1} (b). 
\textcolor{black}{This SC-PCF is a polarization maintaining fiber. 
For this type of fiber, in the general case, phase matching depends significantly on the polarization of the pump at the input to the fiber (see~\cite{petrov2019entropy}). In the context of our work, we obtained the best results when using pump pulses with the polarization direction along the "fast" axis of the fiber.}
In order to analyse a phase matching, we calculated propagation constants $\beta_{s,i}(\omega)$ and corresponding GVD  $\beta^{(2)}(\omega)$  by the effective cladding refractive index model \cite{Knight_1998_model,Koshiba_2005_model}, which is depicted in Fig. \ref{fig:1} (c).
The shown GVD dependence is characterized by two ZDW values of 751 nm and 1230 nm,  which are in agreement with the characteristics of the PCF given by its manufacturer.

\begin{figure}[h!]
\centering
\includegraphics[width=\linewidth]{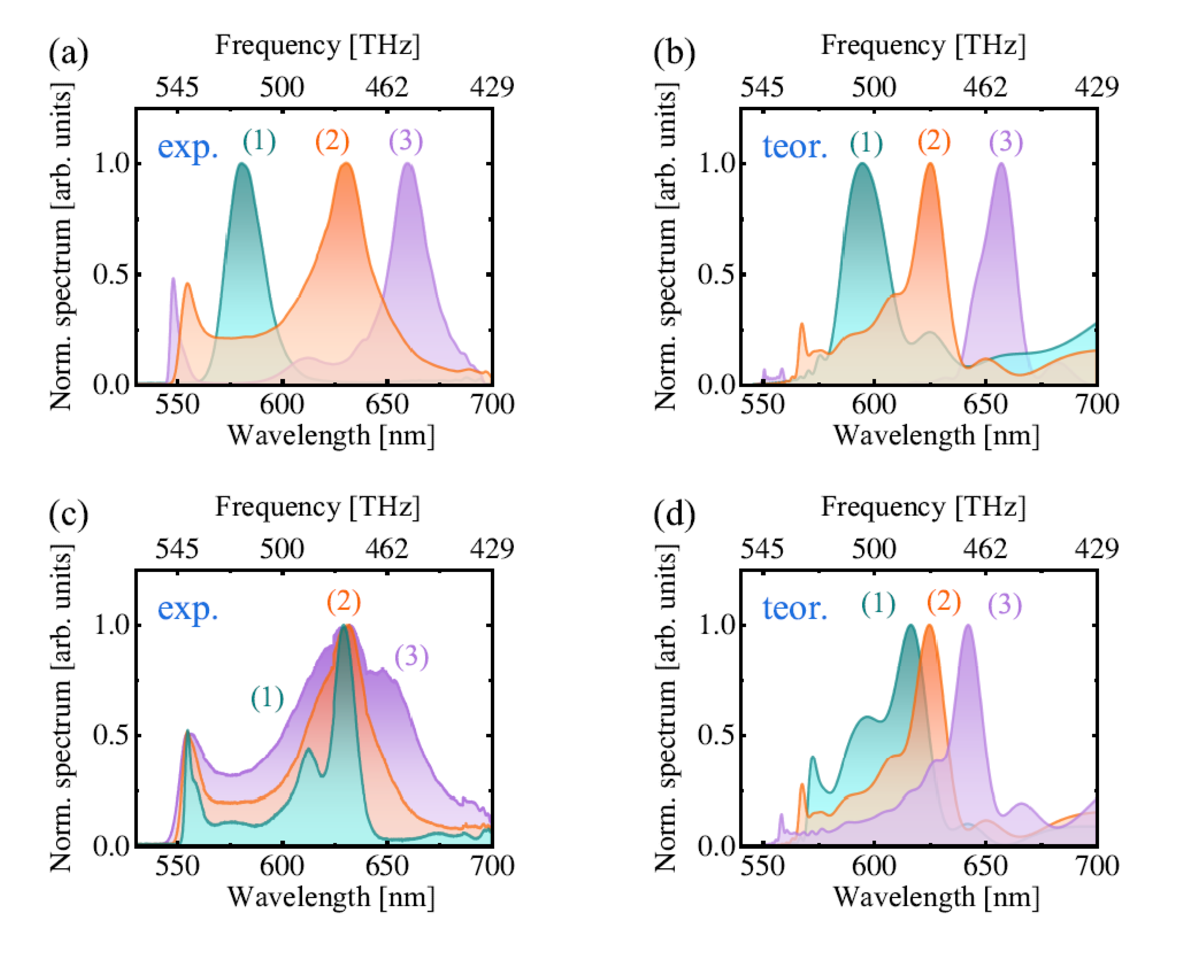}
\caption{Comparison of experimental (a, c) and calculated (b, d) normalized signal photon spectra near the extremum point for different spectral pump parameters: (a) different pump wavelength in experiments ($\Delta\lambda_p\approx$~1.0~nm)
$\lambda_p$=750.2~nm (1), 
$\lambda_p$=750.7~nm (2),
$\lambda_p$=752.0~nm (3);
(b) different pump wavelength in calculations ($\Delta\lambda_p\approx$~0.7~nm)
$\lambda_p$=750.0~nm (1), 
$\lambda_p$=750.7~nm (2),
$\lambda_p$=751.8~nm (3);
(c) different pump spectral wide in experiments
$\Delta\lambda_p$=0.4~nm (1), 
$\Delta\lambda_p$=1.2~nm (2),
$\Delta\lambda_p$=2.2~nm (3);
(d) different pump spectral wide in calculations
$\Delta\lambda_p$=0.5~nm (1), 
$\Delta\lambda_p$=0.9~nm (2),
$\Delta\lambda_p$=2.3~nm (3).} 
\label{fig:2}
\end{figure}

The based on the dispersion phase-matching plot is presented in Fig. \ref{fig:1} (d) as a dependence photon pairs wavelengths from the pump wavelength. 
At a pump wavelength of 750 nm (dashed vertical line) the group velocities of idler and signal photons are matched and the generation of photon pairs with ultrabroad spectral band at  the signal carrier wavelength 600 nm and the idler at 1000 nm is possible. 

First of all, we measured spectra at pump wavelengths around zero dispersion. 
In order to generate the biphoton fields with the largest spectrum width, we accurate tuned a pump wavelength  $\lambda_p$ in the range of 2 nm with fixed half-width $\Delta\lambda_p$=0.7 nm.
Experimental results are demonstrated in Fig. \ref{fig:2}~(a), where we see the sharp photons spectrum changes at  small tuning of the pump wavelength. In this experiment  we  achived the most wide spectrum  at $\lambda_p$=750.7 nm.

Around this pump wavelength, we measured the signal band spectra as a function of the pump half-width $\Delta\lambda_p$. 
The results are presented in Fig. \ref{fig:2}~(a), where we show that the spectra of the generated photons are separated into two spectral bands at small values of the pump width. 
However, as it is seen in Fig. \ref{fig:2} (c), an increase in the spectral width of the pump pulse $\Delta\lambda_p$ to only 2 nm ($\sim$ 1 THz) leads to a considerable broadening of the spectral width of generated photons by more than 50 times (i.e. 100 THz). This effect is achieved by using the carrier frequency of pumping near the point of $\lambda_p$=750.7 nm (see Fig.1 (d)), where  a sharp increase in the spectral width is due to a large steepness of the phase-matching plot.
Similar behavior should be expected from idler lines of the spectrum, based on Eq.~\ref{eq:Eq.1}. 
\textcolor{black}{Numerical calculations of the spectra of signal photons were also performed based on Eq.~\ref{eq:Eq.1}-~\ref{eq:Eq.5} and using the dispersion of our SC-PCF. The calculation results shown in Fig.~\ref{fig:2}~(b,d) demonstrate good agreement with experimental data.}
The total bandwidth of the spectrum (signal and idle bands) minus the Raman scattering band near the pump in this case can reach $\approx$ 200 THz. 

\begin{figure}[h!]
\centering
\includegraphics[width=\linewidth]{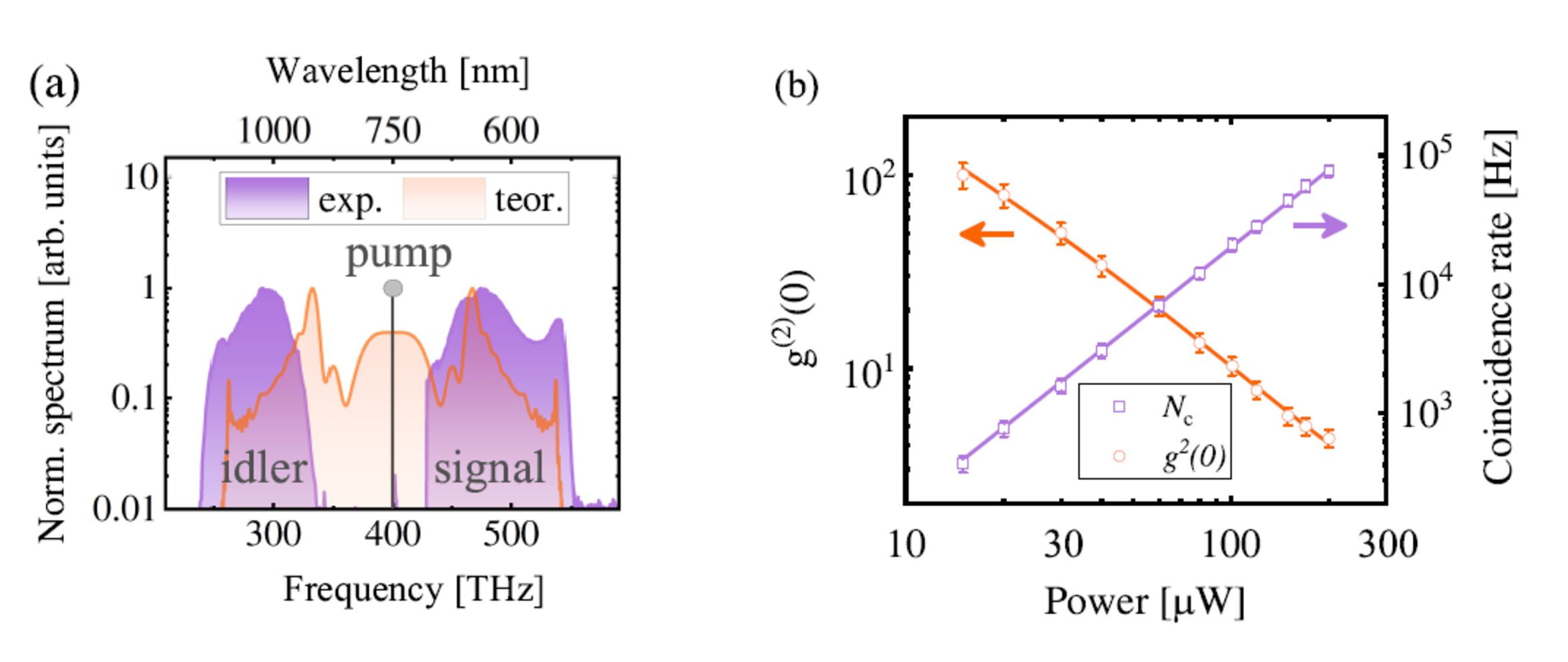}
\caption{(a) Both spectral bands of biphotons (signal and idler bands): experimental (purple) and calculated (orange) spectrum. (b) The second-order correlation function $g^{(2)}(0)$ (left axis, orange circles) and the coincidence rate of the biphoton source (right axis, purple circles) as a function of the pump power. 
Continuous lines describe corresponding approximating functions.
}
\label{fig:4}
\end{figure}

Both spectral bands (signal and idler) are presented in Fig.~\ref{fig:4}(a). 
The idler spectrum was measured by scanning generated beam at monochromator with a coupled infrared InGaAs/InP-SPAD (ID210, IDQuantique).
Analyzing the idler photon spectra, we conclude that  the Raman scattering of pump photons does not significantly affect the biphotons generation in our case. However, the experimental biphoton bandwidth is slightly less and equal to $\approx$~180 THz, which is due to the low detection efficiency at wavelengths below 900 nm of the InGaAs/InP-SPAD. To our knowledge, the obtained biphoton bandwidth is the largest among fiber based biphoton sources.   

The coincidence rate of registration of photon pairs was measured using a Hanbury Brown-Twiss interferometer (see Fig. \ref{fig:1}(a)). 
A single-photon detector based on silicon avalanche diode ($Vis-SPAD$)  with average efficiencies $\eta_s$  of  $65\%$ (in the wavelength range from 540 nm to 680 nm) and a InGaAs/InP-SPAD ($IR-SPAD$) with average efficiencies $\eta_i$  of  $9\%$ (in the wavelength range from 900 nm to 1200 nm) were used for photons detection into signal and idler channels.
The photon counting rates and  coincidence rate were measured using a time to digital converter ($Time~ tagger$) connected to the outputs of these detectors. In order to achieve the necessary spatial separation of the generated photons from pump and Raman scattering photons, high-quality  shortpass and longpass optical filters were installed in the signal and idle channel, respectively. 

We characterized the efficiency of the biphoton source by measuring the second-order correlation function $g^{(2)}(\tau)$ at different pump power. 
The $g^{(2)}(\tau)$  can be obtained as \cite{Chekhova_lopez2021fiber} 

\begin{equation}
{g^{(2)}(\tau)=\frac{N_c(\tau) f_p}{N_s N_i}},
\label{eq:Eq.6}
\end{equation}
where $N_s$ ($N_i$) is the detection rate of signal (idler) photons, $N_c(\tau)$ is the rate of events of consequent signal-idler photons detection with time delay $\tau=t_1-t_2$.
We consider the case when $\tau=0$ and the $N_c$ makes sense of the coincidence rate. 
The experimental results are depicted in Fig.~\ref{fig:4}(b). We show that the values of $g^{(2)}(0)$ decrease with average pump  power $P$ as  $g^{(2)}(0,P)=1+a/(bP^2+cP)$, where coefficients $a,b,c$ are the fitting parameters.
Deviation from the ideal case of biphoton generation (where $g^{(2)}(0,P)\propto 1/P^2$)),  we explain by the presence of a small photon noise manifestation. 
In the experiment we obtained maximum value of $g^{(2)}(0) \approx100$ at low pump power, which demonstrates a nonclassical correlation between signal and idler modes. 
At the same time the values of $N_c$  increase with average pump power as a function $N_c(P)\propto P^2$ in according to the SFWM process. 
We characterise the performance of our fiber-based biphoton source as $g^{(2)}(0) N_c /  \eta_s \eta_i$  which is independent of pump power and in our case achieved the value of $\sim$2,6 MHz. 
This value is 1000 times higher than the value obtained from the data of recent work for HC-PCF based broadband biphoton source \cite{Chekhova_lopez2021fiber}. 

\section{Conclusion}
In conclusion, we have proposed and experimentally demonstrated a high performance fiber-based source of ultra-broadband biphotons with a  record spectral width of the output beam.
We used the phenomenon of spontaneous four-wave mixing in a silica-core PCF with two values of ZDW pumped by femtosecond laser pulses. 
To achieve ultra-broadband spectrum, we carefully set the pump wavelength close to the  short-wavelength ZDW  when phase-matching curve has a a large steepness. 
By using a pulsed pump with spectral bandwidth $\sim$ 1 THz, the maximal spectral width of  generated signal and idler photons achieved $\approx$ 180 THz with high value of performance parameter  $\sim$2,6 MHz and second-order correlation function $\approx100$.
The last two values indicate a low level of optical noise in the output beam, caused by Raman scattering.  

The record spectral width of our biphoton source indicates high prospects for its use in QOCT, 
which will increase its axial resolution up to the submicron scale.
Due to the higher spectral brightness, the demonstrated biphoton source promises to be a powerful tool in two-photon absorption fast spectroscopy of complex molecular systems 
and biological substances, which are characterized by transparency windows in the frequency ranges near 600-900 nm and 1000-1400 nm.
The demonstrated generation of ultra-broadband pairs is convenient for the creation of highly entangled biphotons under femtosecond pumping.
It is worth noting that by choosing a suitable design, we can create PCFs in which the carrier frequency of the steep dispersion can have different values extending from the visible frequency range to the infrared spectral range.
The use of a series of the PCFs will allow for the ultrafast biphoton spectroscopy in a very wide spectral range.

\section{Funding}

This work is carried out with financial support of the Ministry of Education and Science of Russia, Reg. No. NIOKRT 121020400113- 1 and Russian Science Foundation (22-12-00149).

\begin{acknowledgement}

The authors would like to express their deep gratitude to Professor Alexey Zheltikov, who made a great contribution to the creation of this work and the development of the topic of fiber and quantum optics in our research laboratory.

\end{acknowledgement}




\bibliography{Bib_Collection}

\end{document}